\documentstyle[aps,prd,preprint,tighten,floats,epsf]{revtex}
\newlength{\singsp}
\def \MSbar {\vbox{\hrule\kern 1pt\hbox{\rm MS}}}

\begin{document}
%
%*************
%\preprint{\vbox{
\preprint{VAND-TH-98-01}
\bigskip
\bigskip
\bigskip
\bigskip
\bigskip
\bigskip

\title{A Quantum Field Theory Warm Inflation Model}

\bigskip

\author{Arjun Berera}

\medskip 

\address{Department of Physics and Astronomy, Vanderbilt University,
Nashville, TN 37235, USA}

\bigskip

\maketitle

\begin{abstract}
A quantum field theory warm inflation model is presented that solves the
cosmological horizon/flatness problems.  
An interpretation of the model is given
from supersymmetry and superstring theory.
\end{abstract}

\bigskip
\bigskip
\bigskip
\bigskip

To appear in Proceedings COSMOS 98, Monterey, CA,  November 1998

\eject

It has been known for a long time to us \cite{sinf}
(for a review of earlier work see \cite{olive}) and
perhaps a longer time to Nature that inflation is a very attractive
solution to the cosmological puzzles.  Yet despite its simple picture, a
dynamical realization of inflation has proven to be an arduous task.
{}For a long time cosmologists adhered to the notion that a de Sitter
expansion regime would necessitate a rapid depletion of radiation energy
density $\rho_r$ thus creating a supercooled environment during
inflation.  Warm inflation cosmology \cite{wi}
has clarified this misconception
by demonstrating the viability of concurrent radiation
production during an inflationary regime.  
{}Furthermore, the warm inflation picture has a couple of immediate conceptional
advantages.  {}Firstly the dynamics is completely free of questions about
the quantum-to-classical transition.  The scalar inflaton field is in a
well defined classical state, thus immediately justifying the application
of a classical evolution equation.  Also, the fluctuations of the
inflaton, which are the metric perturbations \cite{bf2}, are classical.
Secondly the dynamics underlying warm inflation is based on the best
understood nonequilibrium regime, the state perturbed from thermal
equilibrium.  In this regime, a self-consistent prescription for dynamics
is well defined.  These two points imply the dynamics is free of
conceptual ambiguity, thus permitting a clear road towards a theory.
Notwithstanding, the challenge is to find models that
satisfy the requirements of this prescription.

In this talk, a quantum field theory warm inflation 
model is presented, based on the analysis in \cite{bgr,bgr2}, that solves the
horizon/flatness problems.
The model obtains, from the elementary
dynamics of particle physics, cosmological
scale factor trajectories that begin in a radiation dominated regime,
enter an inflationary regime and then smoothly exit back into a
radiation dominated regime, with nonnegligible radiation
throughout the evolution.

The basic idea of our implementation of warm inflation is quite simple;
a scalar field, which we call the inflaton, interacts with several other 
fields through shifted couplings $g^2(\varphi-M_i)^2 \chi_i^2$ and
$g(\varphi-M_i) {\bar \psi} \psi$ to bosons and fermions repectively.
The mass sites $M_i$ are distributed over some range.
As the inflaton relaxes toward its minimum energy configuration, it will decay
into all fields that are light and coupled to it.  In turn this generates 
an effective viscosity. That this indeed 
happens has been demonstrated in detail in Refs. \cite{bgr,GR}.
In order to satisfy one of the requirements of a successful inflation (60 or so
e-folds), overdamping must be very efficient. The purpose
of distributing the masses $M_i$ is to increase the interval
for $\varphi$ in which light particles emerge through the shifted
couplings.

The basic Lagrangian that we will consider is of a scalar field $\phi$
interacting with $N_M\times  N_{\chi}$ scalar fields $\chi_{jk}$ and
$N_M\times  N_{\psi}$ fermion fields $\psi_{jk}$,

\begin{eqnarray}
\lefteqn{{\cal L} [ \phi, \chi_{jk}, \bar{\psi}_{jk}, \psi_{jk}] = 
\frac{1}{2}
(\partial_\mu \phi)^2 - \frac{m^2}{2}\phi^2 -
\frac{\lambda}{4 !} \phi^4} \nonumber\\
& & + \sum_{j=1}^{N_M} \sum_{k=1}^{N_{\chi}} \left\{
\frac{1}{2} (\partial_\mu \chi_{jk})^2
- \frac{f_{jk}}{4!} \chi_{jk}^4 - \frac{g_{jk}^2}{2}
\left(\phi-M_j\right)^2
\chi_{jk}^2 
\right\}
\nonumber \\
& & +\sum_{j=1}^{N_M} \sum_{k=1}^{N_{\psi}}  
\left\{ i \bar{\psi}_{jk} \not\!\partial \psi_{jk} - h_{jk} ( \phi - M_j ) 
\bar{\psi}_{jk} \psi_{jk} \right\}
\: ,
\label{Nfields}
\end{eqnarray}

\noindent 
where all coupling constants are positive: $\lambda$,
$f_{jk},g_{jk}^2, h_{jk}$ $> 0$. {}For simplicity, we consider in the following
$f_{jk}=f$, $g_{jk}=h_{jk}=g$.
Also, we will set $N_{\psi}=N_{\chi}/4$,
which along with our choice of coupling implies a cancelation
of radiatively generated vacuum energy corrections in the effective
potential \cite{dj}.  We call this kind of model a distributed mass
model (DMM), where the interaction between $\phi$ with the $\chi_{jk}$
and $\psi_{jk}$
fields establishes a mass scale distribution for the $\chi_{jk}$ and
$\psi_{jk}$ fields,
which is determined by the mass parameters $\{M_i\}$. Thus the $\chi_{jk}$
and $\psi_{jk}$ effective field-dependent masses, 
$m_{\chi_{jk}} (\phi,T,\{M\})$ and $m_{\psi_{jk}} (\phi,T,\{M\})$,
respectively, can be constrained
even when $\langle \phi \rangle = \varphi$ is large. The mass
sites are chosen to be $M_i= iT_{M}/g$, where $T_M$ is
a constant that is of order the temperature $T$
during warm inflation.

The above Lagrangian has been realized from an
effective N=1 global SUSY theory with superpotential
\begin{equation}
W(\Phi,\{X_i\})= 4m {\Phi}^2 + \lambda \Phi^3
+\sum_{i=1}^{N_M} \left[4\mu_i X_i^2 + f_i X_i^3 +
\lambda'_i \Phi^2 X_i + \lambda^{''}_i \Phi X_i^2 \right],
\label{superpot}
\end{equation}
which represents an inflaton interacting with the modes of a string.
Here $\Phi$ is a single chiral superfield which represents
the inflaton and $X_i$ $i=1,\ldots,N_M$ are a set of chiral superfields 
that interact with the inflaton.
All the superfields will have their
antichiral superfields $\bar \Phi$, $\{\bar X_i\}$ appearing in kinetic
and Hermetian conjugate (h.c.) terms.  In the
chiral representation the expansion of the superfields in terms of the
Grassmann variable $\theta$ is
$\Phi=\phi + \theta \psi + \theta^2 F$ and 
$X_i= \chi_i + \theta \psi_i + \theta^2 F_i$, $i=1, \ldots , N_M$.
Here $\phi=(\phi_1+ i \phi_2)/\sqrt{2}$ and
$\chi_i = (\chi_1+ i \chi_2)/\sqrt{2}$ are complex scalar fields
as well as $F$ and $\{F_i\}$, and $\psi$ and $\{ \psi_i \}$ are Weyl
spinors.  By definition, the inflaton $\Phi$ characterizes the state
of the vacuum energy through a nonzero amplitude in the bosonic sector
$\langle \phi \rangle \equiv \varphi \neq 0$.

The DM-model is realized for the case
$\mu_i=gM_i/2$, $\lambda_i^{'}=0$, and $\lambda_i^{''}=-2g$, 
for which the masses of the
$\chi_i,\psi_i$ fields are respectively
$m^2_{\chi_i}= g^2(\varphi-M_i)^2 -2gm \varphi -
(3g \lambda/4) \varphi^2$
\label{mchi}
and
$m^2_{\psi_i}= g^2(\varphi-M_i)^2$.
At $\varphi=0$, the masses of the $\chi_i,\psi_i$ pair are
equal,
which is required by supersymmetry.  On the other hand, a nonzero
inflaton field amplitude, $\varphi \neq 0$, implies a soft breaking of
supersymmetry, which in turn permits mass differences.
It has been checked that the soft breaking terms do not cause any problems
for the
results in
\cite{bgr2}.

The hierarchy of mass levels in the above model Eqs. 
(\ref{Nfields}) and (\ref{superpot}) has a
reminiscient similarity to the mass levels of a string. Clearly the above
superpotential, thus the DM-model, captures this basic feature of strings.
Also, since the DM-model can be derived from {}F-term SUSY, i.e. the
superpotential, it is a natural model in the technical sense of
renormalizability.  {}Further  details on matters related to the SUSY origin
of the DM-model and its string interpretation can be found in \cite{bk1}.

The next task is to
derive the effective equation of motion for $\varphi$ that descibes
dissipative dynamics.  The basic idea underlying dissipative dynamics is
very simple.  The decay products of $\phi$ and fields to which it couples
create, in a sense, a viscous fluid which 
in turn acts to slow the motion of $\varphi$.
The 1-loop effective equation of motion for the scalar field $\phi$ is
obtained by setting $\phi = \varphi + \eta$ in Eq. (\ref{Nfields}) and
imposing $\langle \eta \rangle=0$. Then from Weinberg's tadpole method
the 1-loop evolution equation for $\varphi$ (for
homogeneous field) is
\begin{eqnarray}
\lefteqn{\ddot{\varphi} + 3 H \dot{\varphi} +
m^2 \varphi + \frac{\lambda}{6} \varphi^3 +
\frac{\lambda}{2} \varphi \langle \eta^2 \rangle } \nonumber \\
& & + g^2
\sum_i^{N_M} \sum_j^{N_{\chi}} (\varphi-M_i) \langle \chi_{ij}^2 \rangle + 
g \sum_i^{N_M} \sum_j^{N_{\chi}/4} \langle \psi_{ij} \bar{\psi}_{ij} \rangle  = 0 \;.
\label{eqnew}
\end{eqnarray}
\noindent 
In the above, the term $3 H \dot{\varphi} $ describes the
energy red-shift of $\varphi$ due to the expansion of the Universe. This
term comes naturally once we start with a background expanding metric
for Eq. (\ref{Nfields}). In the warm-inflation regime of interest here,
the thermalization condition must hold, which requires the
characteristic time scales (given by the inverse of the decay width) for
the fields in Eq. (\ref{Nfields}) to be faster than the expansion time scale,
$H \gg \Gamma$, 
where $\Gamma$ are decay widths given in \cite{bgr,bgr2}.
In this case, the calculation of the (renormalized) thermal
averages in Eq. (\ref{eqnew}) can be approximated just as in the Minkowski
space-time case.  A systematic perturbative evaluation of the averages 
in the adiabatic, strong dissipative regime was 
presented in \cite{bgr} and re-derived in \cite{yl} with
extension to fermions.  
Based on the systematic perturbative approach, the effective equation of
motion for $\varphi$ is

\begin{equation}
\ddot{\varphi} + V^\prime_{\rm eff} (\varphi,T)
+ \eta (\varphi) \dot{\varphi} = 0 \;,
\label{eqVeff}
\end{equation}

\noindent
where $V^\prime_{\rm eff} (\varphi,T)= \partial V_{\rm
eff} (\varphi,T)/\partial \varphi$ is the field derivative of the
1-loop finite temperature
effective potential, which can be computed by the standard methods.
$\eta (\varphi) \equiv \eta^{\rm B}(\varphi)+\eta^{\rm F}(\varphi)$
is a field dependent dissipation; their explicit expressions are
given in \cite{bgr,yl}.  The above equation of motion is subject to the
thermalization condition and the adiabatic condition,
that the dynamic time-scale for $\varphi$ must be
much larger than the typical collision time-scale ($\sim \Gamma^{-1}$),
$|\varphi/\dot{\varphi}| \gg \Gamma^{-1}$.

The model outlined above has been analyzed in the regime where
$V'(\varphi,T) \approx \lambda \varphi^3/6$ dominates.  To enforce this condition,
further constraints are imposed on the parameter space.  There
is insufficient space to present the complete solution here,
but it can be found in \cite{bgr,bgr2}.  To summaries the results, we
find observationally large e-folds $N_e > 60$ in the regime
$g < 1$, $N \sim 10$, $\varphi/T \sim 10^3$, and $\lambda \sim 10^{-9}$.
In addition, a large number of mass sites $M_i$ are necessary $N_M \sim
10^3$.  The total number of particle fields necessary for the 
dynamics in this regime,
$N N_M \sim 10^4$, is not inconsistent with the particle content
of excited states in string theory \cite{bk1}.

In summary, the model described above has two
appealing features. {}Firstly, since the dynamics is
derived from a first principles treatment of thermal field theory, which
drives inflation from the natural dynamics of a scalar field,
slow roll is a consequence of the dynamics and not an input.  Secondly,
the model offers an
interesting connection to high energy unification, through its relation 
to superstrings.  {}Finally an aside, it is interesting to examine whether
small warm inflations, $N_e \sim 1$, can be implemented as reheating
phases after supercooled inflation.

%\end{references}

%\end{center}
%\begin{verbatim}
\end{document}